\documentclass[acmsmall,screen]{acmart}

\usepackage{booktabs}
\usepackage{tabularx}
\usepackage{array}
\usepackage{multirow}
\usepackage{amsmath}
\usepackage{graphicx}
\usepackage{xcolor}
\usepackage{microtype}
\usepackage{enumitem}
\usepackage{placeins}
\usepackage{float}

\setcopyright{none}
\renewcommand\footnotetextcopyrightpermission[1]{}
\setcitestyle{numbers,sort}
\newcommand{\sys}{\textsc{ObliVul}}

\begin{document}

\title{ObliVul: Alert-Conditioned Safety Obligation Modeling and Bidirectional Counterfactual Validation for Code Vulnerability Detection}

\author{Heyang Tan}
\email{thy15333007817@163.com}
\affiliation{%
  \institution{Taiyuan University of Technology}
  \city{Taiyuan}
  \country{China}
}

\author{Chengxin Gao}
\email{gaochengxin@tyut.edu.cn}
\affiliation{%
  \institution{Taiyuan University of Technology}
  \city{Taiyuan}
  \country{China}
}

\author{Xin Wen}
\email{xwen@tyut.edu.cn}
\affiliation{%
  \institution{Taiyuan University of Technology}
  \city{Taiyuan}
  \country{China}
}

\author{Jiaxin Li}
\email{lijiaxin1926@link.tyut.edu.cn}
\affiliation{%
  \institution{Taiyuan University of Technology}
  \city{Taiyuan}
  \country{China}
}

\author{Rui Cao}
\email{caorui@tyut.edu.cn}
\affiliation{%
  \institution{Taiyuan University of Technology}
  \city{Taiyuan}
  \country{China}
}

\begin{abstract}
In real-world software development, the primary challenge in vulnerability detection is often not finding suspicious code, but identifying which alerts among the large number of candidate alerts produced by static analysis truly warrant attention. Existing learning-based methods mainly identify suspicious patterns at the function or line level, making it difficult to extract complete program evidence centered on an individual alert. Although large language models can infer risk sources, dangerous operations, protection conditions, and state preconditions from local program facts, such semantic information cannot be reliably aligned with specific program nodes, dependency relations, and propagation paths, and is therefore insufficient to verify whether the corresponding safety obligations truly affect the current alert. To address this problem, we propose \sys{}, an alert-conditioned safety obligation modeling and bidirectional counterfactual validation framework for vulnerability detection. For each candidate alert, \sys{} first extracts a Local Evidence Pack (LEP) from the Code Property Graph (CPG) and uses a large language model to recover candidate safety obligations. It then aligns the safety obligations with program nodes, dependency edges, and path scopes to construct a Local Safety Obligation Graph (LSOG). Finally, VAFI aggregates complementary verified alert evidence while suppressing redundant or weaker evidence to produce a function-level vulnerability prediction. Experimental results show that \sys{} effectively distinguishes vulnerable versions from fixed versions and reduces persistent false positives on fixed code. Ablation studies further confirm the necessity of each component for safety obligation recovery, risk response validation, and function-level vulnerability inference.
\end{abstract}



\maketitle
\hypersetup{linkcolor=black}

\section{Introduction}
\label{sec:introduction}

In real-world software development, the foremost challenge in vulnerability detection is often not finding suspicious code, but determining which alerts among a large number truly warrant attention. However, to cover potential risks, static analysis usually needs to conservatively approximate factors such as path reachability and alias relationships, thereby producing large numbers of unverified potential-vulnerability leads. We refer to such leads as \emph{candidate vulnerability alerts}, hereafter abbreviated as ``alerts,'' and define a \emph{safety obligation} (SO) as a security condition on which the validity or invalidity of an alert depends and that can be supported by program evidence and further checked. Alert validation therefore faces three questions: how to extract relevant program evidence around an individual alert, how to recover from this evidence the safety obligations that affect whether the alert is valid, and how to verify that the recovered safety obligations actually act on the current risk.

First, existing methods typically take a complete function, code slice, or statement sequence as the prediction object and output function-level labels or line-level suspiciousness by learning lexical, syntactic, and contextual patterns~\cite{li2018vuldeepecker,li2021sysevr,zhou2019devign}. Predicting a function as vulnerable does not imply that every alert within it is valid, and line-level suspiciousness cannot establish whether an alert has complete support from data dependencies, control dependencies, and path reachability. In particular, when vulnerable and fixed versions share most code structures and dangerous operations, methods that rely on holistic function representations tend to produce the same prediction for both versions~\cite{chakraborty2022arewethere,ding2025primevul}. Second, recovering the safety obligations that affect whether an alert is valid from local program facts remains challenging. Whether an alert is valid depends on whether external input can reach a dangerous operation, whether the relevant branches are reachable, and whether protection or sanitization measures cover the risk path and take effect before the dangerous operation. Although a Code Property Graph (CPG) can uniformly represent abstract syntax, control flow, data dependencies, and call relations, it is difficult for a CPG to directly characterize safety obligations together with their validity, satisfaction status, and scope.

An LLM can infer risk sources, dangerous operations, protection conditions, and state preconditions from local program facts~\cite{sun2024llm4vuln,zhou2024llmreview}, but existing LLM-enhanced methods typically output only vulnerability labels, risk scores, natural-language explanations, or repair suggestions. These outputs are difficult to reliably align with concrete program nodes, dependency edges, and propagation paths, and are also difficult to further check through program analysis. The safety semantics recovered by an LLM must therefore be anchored to concrete program structures and organized as safety obligations with program locations, satisfaction status, scopes, and relational constraints.

Finally, even when safety obligations have been identified and anchored to program structures, model explanations or static correspondences alone remain insufficient to determine whether they genuinely affect the current alert. However, existing studies mostly use vulnerable/fixed pairs for model training, result comparison, or patch identification~\cite{fan2020bigvul,bhandari2021cvefixes,chen2023diversevul} and rarely verify whether changes in safety obligations cause risk responses along the same alert path. More reliable alert validation must not only determine the data and control dependencies among inputs, protection conditions, and dangerous operations, but also perform local interventions constrained by program structures to observe whether risk decreases after missing protection is added and whether risk reappears after existing protection is removed.

To address these problems, we propose \sys{}, an alert-driven bidirectional counterfactual graph-validation framework. It first extracts alert-relevant context from a function-level CPG to form a Local Evidence Pack (LEP). It then uses an LLM to recover alert-related safety semantics, maps the semantics to program nodes, dependency edges, and propagation paths to form safety obligations, and further constructs a Local Safety Obligation Graph (LSOG). Next, it compiles the safety obligations into two types of graph interventions---adding missing protection and removing existing protection---analyzes the resulting risk changes, and extracts minimal supporting evidence. The main contributions of this work are as follows:

\noindent\textbf{1.}\enspace We propose an alert-conditioned local safety-obligation modeling method. Centered on an individual alert, the method extracts relevant program facts from a function-level CPG to form an LEP, uses an LLM to recover alert-related safety obligations, and constructs an LSOG by checking their correspondences with program nodes, dependency edges, and propagation paths.

\noindent\textbf{2.}\enspace We propose a bidirectional counterfactual graph-validation mechanism for safety obligations. By adding missing protection and removing existing protection, it observes risk reduction and risk restoration, respectively, and combines the consistency between the original and counterfactual paths to extract minimal supporting evidence and generate alert-level validation results.

\noindent\textbf{3.}\enspace We propose VAFI, a validation-aware function-level vulnerability inference method. VAFI encodes local program structures, safety obligations, and counterfactual validation results as alert-evidence units, models complementarity and redundancy among different alerts, and coordinates predictions at different levels through alert-level, function-level, and vulnerable/fixed-pair constraints, thereby suppressing the excessive influence of duplicated or weak evidence on the final prediction.

\noindent\textbf{4.}\enspace We conduct a preliminary real-world validation of ObliVul on the Linux kernel, using the trained model to prioritize alerts and combining it with manual review to identify four real vulnerabilities. We then prepare and submit the corresponding patches, three of which have been merged into the official Linux repository.

\section{Related Work}
\label{sec:related-work}
\subsection{Alert-Oriented Static-Analysis-Based Vulnerability Detection}

Alert-oriented static analysis further uses program slicing and path-sensitive analysis to narrow the relevant code scope and locate key statements that affect whether an alert is valid. A Code Property Graph (CPG) unifies abstract syntax, control flow, data dependencies, and call relations into a queryable graph structure, providing a structural foundation for vulnerability-pattern matching, alert-path recovery, and program-evidence extraction.

To address this problem, prior studies have gradually expanded from alert generation to alert prioritization, classification, and automated handling. The Static Analysis Alert Prioritization and Classification Benchmark provides a unified evaluation basis for alert prioritization and classification~\cite{heckman2008benchmark}. LLM4SA traverses program dependencies according to an alert report, extracts relevant code, and uses an LLM to assist alert review~\cite{li2024staticwarnings}. CodeCureAgent organizes code retrieval, alert classification, patch generation, and validation into an agent workflow to identify false positives and handle true alerts~\cite{joos2026codecureagent}.

However, existing methods generally focus on determining whether an alert is valid or generating corresponding explanations and repair suggestions. Although some methods extract program slices, dependency paths, or graph structures, the relevant data dependencies, control dependencies, and protection conditions remain scattered across different statements and paths and have not yet formed a unified safety-obligation representation. Unlike these methods, ObliVul centers on an individual alert, extracts local program evidence from a CPG, and further organizes it into a structured representation for safety-obligation recovery and validation. This alert-centered organization provides a structural basis for aligning subsequent semantic recovery and validation results with the same program locations and paths.
\subsection{LLM-Enhanced Vulnerability Detection}

Early learning-based vulnerability detection methods mainly represented source code as sequences of tokens, statements, or code slices and used neural networks to learn vulnerability patterns~\cite{harzevili2024slr,cheng2021deepwukong,li2021ivdetect}. With the development of pretrained code models, models such as CodeBERT, CodeT5, and UniXCoder learn code representations from large-scale natural-language and source-code corpora, providing richer contextual features for function-level vulnerability detection~\cite{feng2020codebert,wang2021codet5,guo2022unixcoder}. Models represented by LineVul further take complete functions as input, perform function-level prediction using Transformers, and use attention weights to generate line-level suspiciousness rankings~\cite{fu2022linevul}. These methods can capture lexical, syntactic, and local contextual features of code, but data dependencies, control dependencies, and path conditions are usually implicit in sequential representations, making it difficult to directly support the validation of a concrete alert~\cite{steenhoek2024dataflow}.

The first uses code text and external security knowledge as the primary inputs. For example, IRIS combines LLMs with static analysis to supplement semantic judgments in vulnerability discovery, while Vul-RAG retrieves vulnerability knowledge to expand the security-reasoning context of an LLM~\cite{li2024iris,zhou2026vulrag}. The second introduces program graphs or dependency relations into the reasoning process. For example, GRACE uses graph structures and contextual examples to enhance vulnerability detection, while LLMxCPG uses a CPG to guide the model toward vulnerability-related program structures and dependencies~\cite{ding2024grace,lekssays2025llmxcpg}.

However, using a program graph as the input context of an LLM is not equivalent to constraining its output with program structures. If the risk sources, protection conditions, and state preconditions identified by a model cannot be mapped back to concrete program nodes, dependency edges, and propagation paths, the corresponding safety semantics remain difficult to further check through program analysis. To address this problem, ObliVul recovers safety obligations from alert-relevant local evidence and constructs a Local Safety Obligation Graph (LSOG) through program-correspondence checking, transforming LLM-generated safety semantics into structured objects with program anchors and relational constraints.
\subsection{Vulnerable/Fixed Pair Analysis and Bidirectional Counterfactual Validation}

Paired samples comprising vulnerable and fixed versions provide an important basis for analyzing the relationship between local code changes and risk-state changes. Unlike randomly sampled positive and negative examples, vulnerable/fixed pairs share most program context and differ only at local positions such as boundary checks, branch conditions, sanitization operations, or state constraints. PrimeVul constructs a more realistic vulnerability-detection benchmark through strict deduplication and temporal splitting while preserving the correspondence between vulnerable and fixed functions~\cite{ding2025primevul}; subsequent realistic-dataset studies further examine evaluation design and fixed-version reliability~\cite{lecong2024improved,chakraborty2024revisiting}.

Comparing a model's predictions for vulnerable and fixed versions can determine whether the model responds to a risk change introduced by a fix, but cannot explain which edit in the patch causes the risk reversal. When a fix changes multiple statements, conditions, or dependency relations simultaneously, the prediction difference may also arise from contextual changes unrelated to the current alert. Patch analysis can locate code modifications but cannot determine whether newly added protection covers the current dangerous path or genuinely constitutes an effective safety obligation for the alert.

ObliVul uses safety obligations in the LSOG as local program-graph intervention objects and examines the relationship between safety obligations and alert risk in two directions: adding a missing obligation in the vulnerable context to observe whether the dangerous path is suppressed, and removing existing protection in the fixed context to observe whether risk reappears. The two interventions provide risk-reduction and risk-restoration evidence, respectively, and path consistency is used to extract minimal supporting evidence. This process validates structural correspondence and risk responses on the local program graph rather than providing a strict causal proof of complete program behavior; the resulting validation results serve as alert evidence and are aggregated by VAFI into a function-level vulnerability conclusion.

\FloatBarrier

\section{Approach}
\label{sec:method}

\subsection{Framework Overview}
\begin{figure}[H]
\centering
\includegraphics[width=\linewidth]{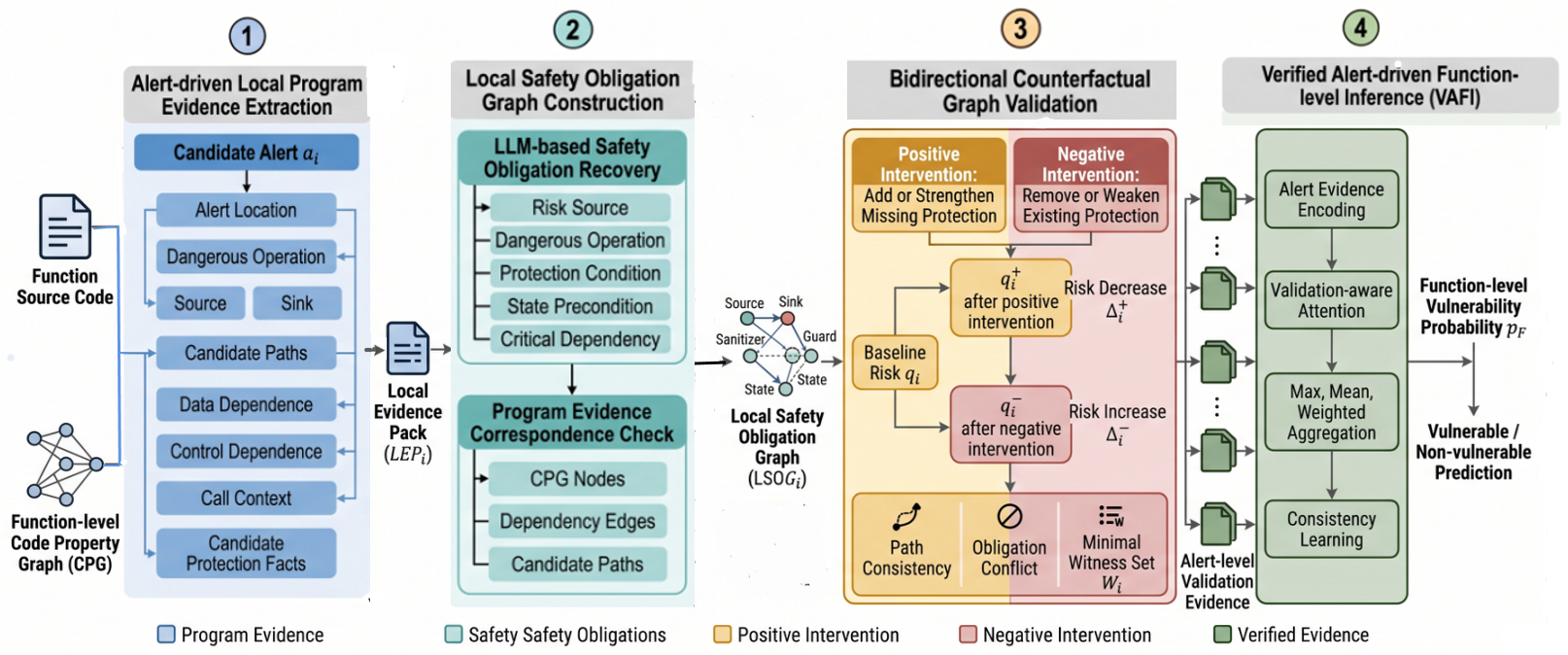}
\caption{Overview of the alert-driven bidirectional counterfactual vulnerability-validation framework.}
\Description{A four-stage pipeline consisting of alert-driven local program-evidence extraction, local safety-obligation graph construction, bidirectional counterfactual graph validation, and validation-aware function-level vulnerability inference.}
\label{fig:framework-overview}
\end{figure}

This paper proposes ObliVul, an alert-driven bidirectional counterfactual vulnerability-validation framework. It sequentially executes four modules: alert-driven local program-evidence extraction, local safety-obligation graph construction, bidirectional counterfactual validation, and validation-aware function-level vulnerability inference. The overall workflow is shown in Figure~\ref{fig:framework-overview}.

\paragraph{(1) Alert-driven local program-evidence extraction.} This module parses the function under analysis into a Code Property Graph (CPG) and generates alerts using static-analysis rules. Centered on the report location, dangerous operation, and candidate paths of an alert, it then extracts the relevant data propagation, control constraints, call relations, and protection conditions to form a Local Evidence Pack (LEP). Subsequent analysis is thereby restricted to the program entities and dependencies that affect the current alert.

\paragraph{(2) Local safety-obligation graph construction.} This module uses an LLM to recover risk sources, dangerous operations, protection conditions, state preconditions, and critical dependencies from alert-level local evidence and determines whether candidate safety obligations are satisfied, missing, or ineffective. It then checks their correspondence with CPG nodes, dependency edges, and propagation paths, filters content unsupported by program evidence, and constructs a Local Safety Obligation Graph (LSOG).

\paragraph{(3) Bidirectional counterfactual graph validation.} This module compiles the safety obligations in an LSOG into two types of local CPG interventions. It adds or strengthens missing or insufficient protection to check whether risk decreases, and removes or weakens existing protection to check whether risk reappears. By comparing path risk before and after intervention, it validates the risk-response relationship between a safety obligation and the current alert and extracts alert-level validation evidence.

\paragraph{(4) Validation-aware function-level vulnerability inference.} This module applies Verified Alert-driven Function-level Inference (VAFI), which encodes the local CPG, LSOG, and counterfactual validation results as alert-evidence units, models complementarity and redundancy among different alerts, aggregates alert evidence within a function, and outputs a function-level risk score and vulnerability decision.

\subsection{Alert-Driven Local Program-Evidence Extraction}
This module first parses the function under analysis into a Code Property Graph (CPG) and executes predefined security checks. A candidate alert is generated when a dangerous operation is detected as potentially risky. For a function $F$ and its Code Property Graph $G_F$, the alert set is

\begin{equation}
A(F)=SA(G_F)=\{a_i\}_{i=1}^{n},
\tag{3-1}
\end{equation}

where $SA(\cdot)$ denotes the static-analysis process and $a_i$ denotes the $i$-th alert. Each alert contains a rule or vulnerability category, report location, dangerous operation, candidate propagation path, and related program context. The report location and dangerous operation determine the center of local analysis, the rule category indicates the possible type of safety obligation, and the candidate path limits the initial evidence scope. An alert represents only a potential risk pattern and is not equivalent to a confirmed vulnerability.

To reduce interference from program information unrelated to the current alert, this paper performs a relation-driven local projection on the CPG conditioned on alert $a_i$. The process first locates the CPG nodes corresponding to the report location and dangerous operation. It then follows the candidate paths to recover potential data sources, propagation variables, and interprocedural call relations, and supplements the control conditions that affect path reachability and the boundary checks, sanitization operations, and state constraints that may suppress risk. The local program graph corresponding to the alert is defined as

\begin{equation}
G_i^{\mathrm{local}}=\operatorname{Proj}_{\mathcal{R}}(G_F;a_i),
\tag{3-2}
\end{equation}

where $\operatorname{Proj}_{\mathcal{R}}$ denotes alert-conditioned local projection. The relation set $\mathcal{R}$ includes data dependencies, control dependencies, call relations, and candidate protection relations. This stage extracts only candidate protection facts. Whether they constitute effective safety obligations is further determined in Section~3.3. Interprocedural analysis retains only call sites, parameter passing, return-value propagation, and restricted callee context directly related to the current alert, without expanding irrelevant complete call subgraphs.

To control evidence size, projection stops upon reaching a candidate data source, a predefined expansion depth, or an interprocedural boundary. When multiple candidate paths exist, only relevant paths satisfying length and number limits are retained; the specific thresholds are given in the experimental settings.

Based on the local program graph, the program facts related to alert $a_i$ are organized as a Local Evidence Pack (LEP):

\begin{equation}
LEP_i=\left(G_i^{\mathrm{local}},X_i^{\mathrm{data}},X_i^{\mathrm{ctrl}},X_i^{\mathrm{call}},X_i^{\mathrm{protect}}\right),
\tag{3-3}
\end{equation}

where $G_i^{\mathrm{local}}$ retains the local CPG structure corresponding to the alert. $X_i^{\mathrm{data}}$, $X_i^{\mathrm{ctrl}}$, $X_i^{\mathrm{call}}$, and $X_i^{\mathrm{protect}}$ are normalized relation views extracted from the local graph and represent data propagation, control constraints, interprocedural transfer, and candidate protection coverage, respectively. These views provide explicit and parseable program facts to the LLM and do not constitute additional evidence independent of the CPG.

\subsection{Local Safety-Obligation Graph Construction}
The Local Evidence Pack $LEP_i$ extracted in Section~3.2 describes alert-related data propagation, control dependencies, call relations, and candidate protection facts, but does not yet specify the safety obligations on which the presence or suppression of the current alert depends. This paper therefore uses an LLM to induce alert-related local safety obligations from $LEP_i$ and organizes them into a Local Safety Obligation Graph (LSOG).

\subsubsection{Alert-Conditioned Safety-Obligation Induction}
For alert $a_i$, the alert information and Local Evidence Pack $LEP_i$ are jointly provided to an LLM to recover security entities, program roles, and risk conditions associated with the current alert. The LLM does not directly output a conclusion on whether the alert is valid. Instead, it generates a candidate safety-obligation set:

\begin{equation}
\widetilde{O}_i=LLM(a_i,LEP_i,P_i^{\mathrm{ctx}}),
\tag{3-4}
\end{equation}

where $\widetilde{O}_i$ denotes the candidate safety-obligation set and $P_i^{\mathrm{ctx}}$ denotes optional project-context priors.

Candidate safety obligations mainly include risk sources, dangerous operations, required protection, ineffective protection, state preconditions, and critical dependencies. Compared with directly predicting vulnerability labels, alert-conditioned induction decomposes risk judgment into checkable safety obligations and thereby describes how risk forms, why protection fails, and which alert path the relevant conditions affect.

Because LLM-generated obligations remain semantic hypotheses formed from local context, they may contain incorrectly located program entities, unsupported dependencies, or inaccurate scope judgments. To prevent semantic induction from becoming detached from program structure, this paper further checks the correspondence between candidate obligations and nodes, dependency edges, and propagation paths in the local CPG:

\begin{equation}
O_i=\operatorname{Align}_{G_i^{\mathrm{local}}}(\widetilde{O}_i),
\tag{3-5}
\end{equation}

where $\operatorname{Align}_{G_i^{\mathrm{local}}}$ denotes program-correspondence checking based on the local CPG and $O_i$ denotes the checked safety-obligation set. The checking process maps variables, operations, and conditions involved in a candidate obligation to the corresponding CPG nodes or paths, verifies their data, control, or call dependencies, and determines whether the obligation scope covers the dangerous path corresponding to the current alert.

According to the checking results, candidate obligations are divided into four cases: accepted, partially supported, conflicting, and rejected.

For a missing-protection obligation, the system establishes an expected program anchor from the relevant data object, dangerous path, and expected protection location to represent where and over what scope the protection should act. Such an obligation is marked as missing and becomes a candidate for subsequent forward graph intervention. To provide a unified representation of the results after program-correspondence checking, each safety obligation is represented as

\begin{equation}
o_j=(type_j,anchor_j,condition_j,status_j,polarity_j,scope_j,confidence_j),
\tag{3-6}
\end{equation}

where $type_j$ denotes the obligation type, including risk source, dangerous operation, protection condition, state precondition, and critical dependency. $anchor_j$ denotes the corresponding CPG node, dependency edge, or propagation path; for missing protection, it records the expected operation location. $condition_j$ describes the specific condition that the obligation requires or checks. $scope_j$ limits the propagation path, control region, or call context to which the condition applies. $polarity_j$ indicates whether the obligation promotes or suppresses risk.

The main values of $status_j$ are $present$, $missing$, $partial$, and $conflict$. $present$ means that the corresponding program fact exists and reverse validation can be performed. $missing$ means that the current path lacks the corresponding protection and forward validation can be performed. $partial$ means that the obligation is supported by only part of the program evidence. $conflict$ means that the obligation is inconsistent with program facts and is retained only as conflicting evidence.

For ineffective protection, its existence can be represented by $present$, while the specific reason for its ineffectiveness is jointly described by the obligation type, condition, scope, and program correspondence. $confidence_j$ is not a subjective probability directly produced by the LLM. Instead, it is a structured indicator that jointly reflects semantic clarity, program-anchor alignment, path coverage, dependency support, and obligation conflict.

\subsubsection{Local Safety-Obligation Graph Construction}
After program-correspondence checking, the relationships among different obligations along the current alert path must also be represented. An individual safety obligation cannot completely express the joint effects of risk sources, dangerous operations, control constraints, and protection measures. The filtered safety obligations are therefore organized into an alert-conditioned Local Safety Obligation Graph according to the data propagation, control dependencies, and protection-coverage relations in the local CPG:

\begin{equation}
LSOG_i=\left(O_i,R_i^{\mathrm{data}},R_i^{\mathrm{ctrl}},R_i^{\mathrm{protect}}\right),
\tag{3-7}
\end{equation}

where $O_i$ denotes the safety-obligation set after program-correspondence checking; $R_i^{\mathrm{data}}$ denotes data-propagation relations among risk sources, propagation objects, and dangerous operations; $R_i^{\mathrm{ctrl}}$ denotes the control relations of branch conditions and state checks over dangerous paths; and $R_i^{\mathrm{protect}}$ denotes the protection-coverage relations of boundary checks, sanitization operations, or state constraints over target dangerous paths. These relations are jointly determined by data dependencies, control dependencies, and call relations in the local CPG and by the candidate protection facts extracted in Section~3.2.

In an LSOG, nodes represent risk entities, safety obligations, and satisfaction states associated with the current alert, while edges represent data propagation, control constraints, and protection coverage among nodes. Entities or conditions that can be mapped to the original program retain their corresponding CPG program anchors. For a missing safety obligation, the LSOG records the data object, dangerous path, and operation location to which its expected constraint should apply and marks its state as $missing$. For partially supported or conflicting obligations, the LSOG retains their states and conflict information to distinguish effective protection, weakly supported protection, and inconsistent obligations.

\subsection{Bidirectional Counterfactual Graph Validation}
In Section~3.3, the LSOG organizes alert-related risk entities, protection conditions, and their relations as structured safety obligations. These obligations, however, remain candidate safety semantics formed from local program evidence and cannot directly prove that they affect the current alert. This paper therefore compiles safety obligations in an LSOG into local graph interventions and reanalyzes risk changes on path states derived from the local CPG to test whether a safety obligation is consistent with the risk change of the current alert.

\subsubsection{LSOG-Based Safety-Obligation Compilation and Bidirectional Counterfactual Validation}
For the Local Safety Obligation Graph $LSOG_i$ associated with alert $a_i$, this paper compiles safety obligations into forward and reverse local graph interventions according to their types, states, program anchors, and scopes.

Here, $\mathcal{T}_i^{+}$ denotes forward intervention. It mainly targets missing or ineffective protection conditions and simulates a protected dangerous path by adding or strengthening protection. $\mathcal{T}_i^{-}$ denotes reverse intervention. It mainly targets protection conditions already present on the current path and simulates a path state after protection failure by removing or weakening the corresponding constraints. A protection obligation with state $missing$ is compiled into a forward intervention, while an obligation with state $present$ is compiled into a reverse intervention. An obligation without a clear program anchor or path support does not directly participate in graph intervention.

To compare risk changes produced by forward and reverse interventions under a unified baseline, this paper establishes the local path state before intervention. This state consists of the source, sink, candidate propagation paths, control conditions, and protection facts in the local CPG. Candidate-path risk is then reanalyzed and aggregated to obtain the baseline risk of the current alert:

\begin{equation}
q_i=\operatorname{Risk}(G_i^{\mathrm{local}},LSOG_i),
\tag{3-8}
\end{equation}

where $\operatorname{Risk}(\cdot)$ denotes local path-risk reanalysis and $q_i$ denotes alert risk before graph intervention.

For a safety obligation that is missing from the alert path or does not operate effectively, forward validation applies $\mathcal{T}_i^{+}$ to the original local path state and reanalyzes risk after adding or strengthening the corresponding protection:

\begin{equation}
q_i^{+}=\operatorname{Risk}\left(\mathcal{T}_i^{+}(G_i^{\mathrm{local}}),LSOG_i\right),
\qquad
\Delta_i^{+}=q_i-q_i^{+}.
\tag{3-9}
\end{equation}

Here, $q_i^{+}$ denotes alert risk after forward intervention and $\Delta_i^{+}$ denotes its reduction relative to the baseline risk $q_i$. A large $\Delta_i^{+}$ indicates that the candidate safety obligation can cut off or weaken the current dangerous path.

For a protection condition already present on the local path, reverse validation applies $\mathcal{T}_i^{-}$ to the current local path state and reanalyzes risk after removing or weakening the corresponding protection:

\begin{equation}
q_i^{-}=\operatorname{Risk}\left(\mathcal{T}_i^{-}(G_i^{\mathrm{local}}),LSOG_i\right),
\qquad
\Delta_i^{-}=q_i^{-}-q_i.
\tag{3-10}
\end{equation}

Here, $q_i^{-}$ denotes alert risk after reverse intervention and $\Delta_i^{-}$ denotes its increase relative to the baseline risk $q_i$. A large $\Delta_i^{-}$ indicates that the corresponding protection participates in restricting the current dangerous path.

For vulnerable/fixed paired samples, forward validation can correspond to adding to the vulnerable version the safety obligation implied by the fix, while reverse validation can correspond to removing the corresponding protection from the fixed version.

Based on these results, bidirectional risk changes, consistency between the validation and original alert paths, and the degree of obligation conflict are combined into an alert-level validation score:

\begin{equation}
v_i=\alpha\Delta_i^{+}+\beta\Delta_i^{-}+\gamma A_i-\eta C_i.
\tag{3-11}
\end{equation}

Here, $A_i$ denotes the degree of consistency between the intervention path and the original alert path, and $C_i$ denotes the degree of conflict among obligations or between obligations and local program evidence. $\alpha$, $\beta$, $\gamma$, and $\eta$ control the contributions of the validation signals to the alert-level score.

For an alert containing multiple intervenable safety obligations, this paper progressively selects intervention items that still preserve the current risk response and extracts minimal supporting evidence. This evidence records the safety obligations, operation locations, and related paths that cause the main risk change, reducing the influence of repeated or irrelevant conditions on the validation result. Each alert ultimately forms an alert-level validation result comprising baseline risk, forward and reverse risk changes, path consistency, obligation conflict, and minimal supporting evidence, which is used as input to VAFI for function-level aggregation.

\subsubsection{Alert Evidence Flow in a PrimeVul Paired Case}
To illustrate how a candidate alert is converted into safety obligations and obtains validation evidence through local program structure and bidirectional counterfactual interventions, we use the PrimeVul paired sample CVE-2021-37681 as an example.

\begin{figure}[t]
\centering
\includegraphics[width=0.90\linewidth]{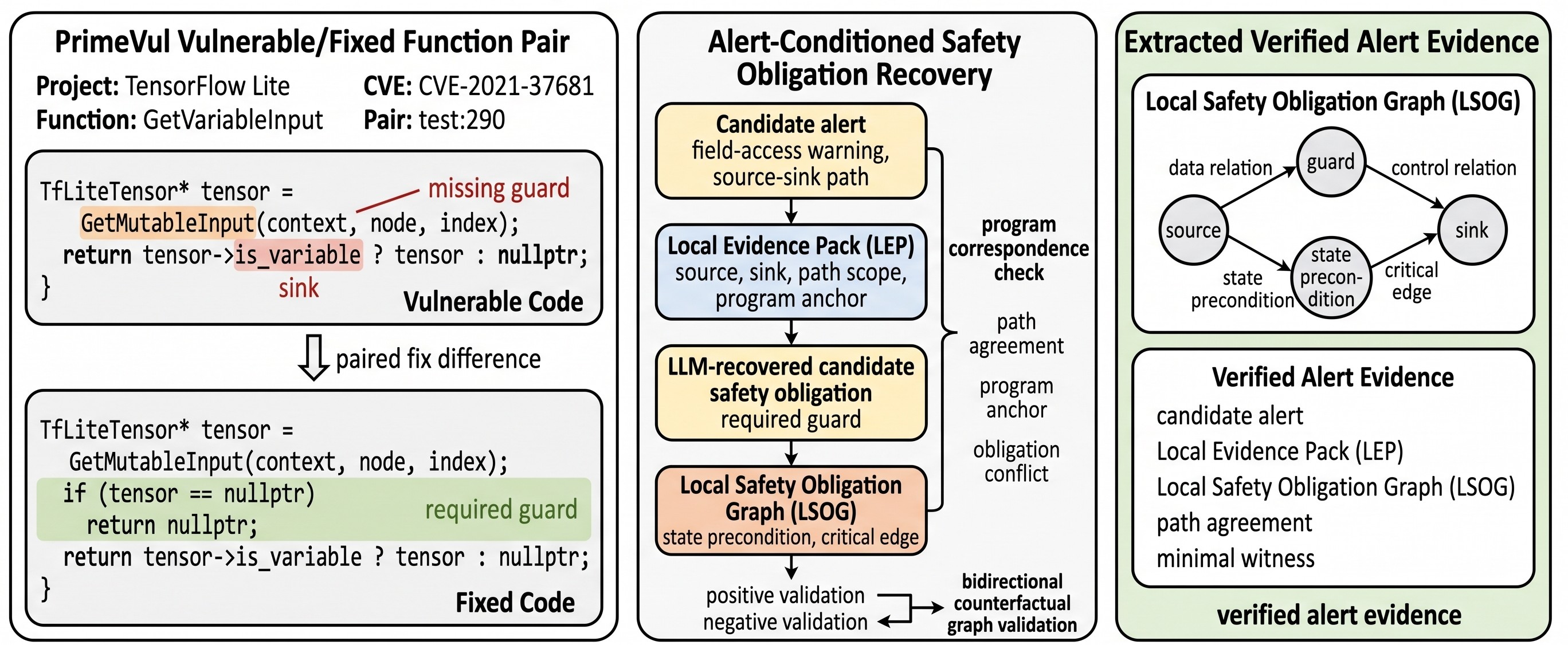}
\caption{CVE-2021-37681: Alert-evidence flow from vulnerability-semantic recovery to bidirectional counterfactual graph validation.}
\Description{A PrimeVul vulnerable/fixed pair illustrating candidate-alert identification, safety-obligation recovery, program-correspondence checking, bidirectional counterfactual validation, and minimal-witness extraction.}
\label{fig:cve-alert-flow}
\end{figure}

In this case, the candidate alert uses tensor->is\_variable as the sink and associates it with its source, source--sink path, path scope, and program anchor through the Local Evidence Pack (LEP). After program-correspondence checking, the safety obligation is represented in the Local Safety Obligation Graph (LSOG). Positive validation then adds the missing guard to the vulnerable version, reducing local path risk; negative validation removes the protection from the fixed version, causing risk to increase again. Both directions show that the guard genuinely restricts the original alert path. The system further extracts a minimal witness containing the required guard, state precondition, and critical path and organizes it together with the alert path and counterfactual risk response as verified alert evidence for subsequent function-level inference by VAFI. Figure~\ref{fig:cve-alert-flow} shows the complete process from candidate-alert identification and safety-obligation recovery through program-correspondence checking, bidirectional counterfactual validation, and minimal-evidence extraction.

\subsection{Verified Alert-Driven Function-Level Inference}
Alert-level validation indicates the degree to which an individual alert is supported by program structure and risk response, but determining whether a function contains a vulnerability requires jointly considering multiple alerts within the function. This paper therefore proposes Verified Alert-driven Function-level Inference (VAFI), which encodes local program evidence, safety obligations, and counterfactual validation results as alert-evidence units and aggregates them to obtain a function-level vulnerability decision.

For the $i$-th alert in function $F$, VAFI encodes its local CPG features, LSOG features, alert risk, bidirectional counterfactual validation results, and minimal supporting evidence into a unified representation:

\begin{equation}
\boldsymbol{h}_i=Encoder\left(\boldsymbol{g}_i^{\mathrm{local}},\boldsymbol{o}_i,\boldsymbol{r}_i,\boldsymbol{c}_i,\boldsymbol{m}_i\right).
\tag{3-12}
\end{equation}

Here, $\boldsymbol{g}_i^{\mathrm{local}}$, $\boldsymbol{o}_i$, $\boldsymbol{r}_i$, $\boldsymbol{c}_i$, and $\boldsymbol{m}_i$ denote the local CPG structure, safety obligations in the LSOG, original alert risk, counterfactual validation results, and minimal supporting evidence, respectively.

VAFI separately estimates the strength of alert risk evidence and the reliability of its security semantics and calculates its weight in function-level aggregation:

\begin{equation}
e_i=AlertHead(\boldsymbol{h}_i),\qquad
s_i=SemanticHead(\boldsymbol{h}_i),\qquad
\omega_i=
\frac{\exp\left(Attn(\boldsymbol{h}_i,e_i,s_i)/\tau\right)}
{\sum_{a_j\in A(F)}\exp\left(Attn(\boldsymbol{h}_j,e_j,s_j)/\tau\right)}.
\tag{3-13}
\end{equation}

Here, $e_i$ denotes the strength with which the alert supports the function-level vulnerability conclusion, $s_i$ denotes the degree of consistency between the security semantics and program evidence, $\omega_i$ denotes the aggregation weight, and $\tau$ is the attention-temperature parameter.

VAFI assigns alert weights through a validation-aware attention mechanism and combines the strongest alert, the alert mean, and the weighted evidence representation. This allows valid alerts from different paths to jointly support the function-level decision while preventing duplicate alerts from being amplified in direct proportion to their number. The function-level vulnerability probability is

\begin{equation}
p_F=\sigma\left(FunctionHead\left(\sum_{a_i\in A(F)}\omega_i\boldsymbol{h}_i,\,\boldsymbol{h}_{max},\,\boldsymbol{h}_{mean},\,z_F\right)\right).
\tag{3-14}
\end{equation}

Here, $\boldsymbol{h}_{max}$, $\boldsymbol{h}_{mean}$, and $z_F$ denote the strongest alert, the alert mean, and function-level anchoring information, respectively. The function-level vulnerability decision is obtained by comparing $p_F$ with a decision threshold determined on the validation set.

VAFI is trained through joint learning. $\mathcal{L}_{func}$ is the function-level classification loss. $\mathcal{L}_{sem}$ and $\mathcal{L}_{alert}$ constrain the consistency of security semantics and alert evidence, respectively. $\mathcal{L}_{rank}$ and $\mathcal{L}_{pair}$ constrain alert risk and function-level risk across vulnerable/fixed versions. $\mathcal{L}_{cons}$ and $\mathcal{L}_{anchor}$ constrain the function prediction to remain consistent with aggregated semantics, program anchors, and path coverage. The overall optimization objective is

\begin{equation}
\mathcal{L}=\mathcal{L}_{func}
+\lambda_{sem}\mathcal{L}_{sem}
+\lambda_{alert}\mathcal{L}_{alert}
+\lambda_{rank}\mathcal{L}_{rank}
+\lambda_{cons}\mathcal{L}_{cons}
+\lambda_{anchor}\mathcal{L}_{anchor}
+\lambda_{pair}\mathcal{L}_{pair}.
\tag{3-15}
\end{equation}

The loss weights are determined on the validation set. Vulnerable/fixed function pairs are used only during training. At test time, the fixed version or repair code corresponding to a function under test is not used. VAFI independently aggregates the candidate alerts of each function. For a function without alerts, an empty-alert mask and the function-level structural representation $z_F$ are used for prediction.

\FloatBarrier
\section{Experimental Evaluation}
\label{sec:evaluation}

\newcommand{\evaltablesetup}{\fontsize{8.2pt}{9.4pt}\selectfont\setlength{\tabcolsep}{1.3pt}\renewcommand{\arraystretch}{1.00}}

\subsection{Experimental Setup}
\subsubsection{Dataset and Experimental Protocol}
This paper uses the pair-function portion of the publicly available PrimeVul dataset, hereafter referred to as PrimeVul pair-function, as the experimental dataset~\cite{ding2025primevul}. PrimeVul pair-function is constructed from C/C++ vulnerabilities in real-world software projects. Its samples originate from public vulnerability reports and their corresponding fixing commits, and the correspondence between vulnerable and fixed functions is preserved. The dataset can therefore reflect how security factors such as boundary checks, data sanitization, state constraints, and control conditions affect vulnerability states.

Each pair consists of corresponding functions before and after the same vulnerability-fixing commit. The function before the fix is labeled as the vulnerable version, and the function after the fix is labeled as the fixed version. Because both versions come from the same fixing context, their function semantics, calling context, and local program structures are highly comparable. The dataset can therefore be used both to determine whether an individual function is vulnerable and to analyze the risk change between vulnerable and fixed versions after safety obligations are added or removed. To further evaluate ObliVul's cross-dataset generalization capability in the Linux kernel vulnerability setting, we additionally use LinuxVul, the official function-pair dataset released by Vul-RAG~\cite{zhou2026vulrag}. LinuxVul contains Linux kernel vulnerability functions collected from vulnerability-fixing commits and their corresponding fixed functions. The pre-fix vulnerable functions are labeled 1, and the corresponding post-fix functions are labeled 0. The LinuxVul training set is used for dataset adaptation and classifier training, the validation set is used to determine model selection, fusion weights, and thresholds, and the test set is used only for final evaluation; the partition is CVE-disjoint, as shown in Table~\ref{tab:dataset-statistics}.

\begin{table}[!htbp]
\captionsetup{skip=1pt}
\caption{Dataset partition used in this paper.}
\label{tab:dataset-statistics}
\centering
\evaltablesetup
\begin{tabularx}{\linewidth}{@{}X>{\centering\arraybackslash}p{0.14\linewidth}>{\centering\arraybackslash}p{0.14\linewidth}>{\centering\arraybackslash}p{0.09\linewidth}@{}}
\toprule
Dataset size & Training & Validation & Test \\
\midrule
PrimeVul original paired samples & 3,789 & 480 & 435 \\
PrimeVul valid paired samples after preprocessing & 3,595 & 461 & 422 \\
LinuxVul original paired samples & --- & --- & 586 \\
LinuxVul valid pairs after preprocessing and CVE-disjoint splitting & \textbf{384} & \textbf{97} & \textbf{97} \\
\bottomrule
\end{tabularx}
\end{table}

\subsubsection{Evaluation Metrics}
Because ObliVul must both perform binary classification of whether an individual function is vulnerable and examine state differences between vulnerable and fixed versions, this paper evaluates model performance at two levels: function-level classification and paired-state diagnosis.

\paragraph{(A) Function-level classification metrics.} We use Accuracy, Precision, Recall, FPR, and F1-score to report function-level classification performance.

\paragraph{(B) Paired-state diagnosis metrics.} Paired-state diagnosis uses P-C, P-V, P-B, P-R, and VPS. P-C denotes that the vulnerable and fixed versions are both correctly distinguished. P-V denotes that both versions are predicted as vulnerable. P-B denotes that both versions are predicted as non-vulnerable. P-R denotes that the prediction direction is opposite to the true version relationship. VPS summarizes the model's effective ability to distinguish vulnerable and fixed versions; a higher value indicates that the model better identifies vulnerable versions while reducing persistent alerts on fixed versions.

\subsection{Comparison Experiments}
This section compares ObliVul with existing methods on the PrimeVul pair-function test set. Table~\ref{tab:main-function} presents the function-level classification results, and Table~\ref{tab:main-pair} presents the paired-state diagnosis results.

\begin{table}[!htbp]
\captionsetup{skip=1pt}
\caption{Function-level results on PrimeVul-paired data.}
\label{tab:main-function}
\centering
\evaltablesetup
\begin{tabular}{@{}llccccc@{}}
\toprule
Method & Dataset & Acc$\uparrow$ & Pre$\uparrow$ & Recall$\uparrow$ & FPR$\downarrow$ & F1$\uparrow$ \\
\midrule
CodeBERT~\cite{feng2020codebert} & PrimeVul & 0.5000 & 0.5000 & 1.0000 & 1.0000 & 0.6667 \\
CodeT5~\cite{wang2021codet5} & PrimeVul & 0.5000 & 0.5000 & 0.9976 & 0.9976 & 0.6661 \\
UniXCoder~\cite{guo2022unixcoder} & PrimeVul & 0.5391 & 0.5409 & 0.5166 & 0.4384 & 0.5285 \\
LineVul~\cite{fu2022linevul} & PrimeVul & 0.6540 & 0.7539 & 0.4573 & 0.1493 & 0.5693 \\
Ding et al. CoT$^{*}$~\cite{ding2025primevul,wei2022cot} & PrimeVul & 0.5126 & 0.5355 & 0.1908 & 0.1655 & 0.2814 \\
GPTLens$^{*}$~\cite{hu2023gptlens} & PrimeVul & 0.5184 & 0.5144 & 0.6552 & 0.6184 & 0.5763 \\
VulTrial$^{*}$~\cite{widyasari2026vultrial} & PrimeVul & 0.5356 & 0.5317 & 0.5977 & 0.5264 & 0.5628 \\
ConColl$^{*}$~\cite{tsai2025concoll} & PrimeVul & 0.5126 & 0.5078 & 0.8253 & 0.8000 & 0.6287 \\
VulTriage$^{*}$~\cite{tang2026vultriage} & PrimeVul & 0.6356 & 0.6000 & 0.8138 & 0.5425 & 0.6907 \\
\textbf{ObliVul} & \textbf{PrimeVul} & \textbf{0.7358} & \textbf{0.6846} & \textbf{0.8744} & \textbf{0.4028} & \textbf{0.7680} \\
\textbf{ObliVul} & \textbf{LinuxVul} & \textbf{0.8969} & \textbf{0.8889} & \textbf{0.9072} & \textbf{0.1134} & \textbf{0.8980} \\
\bottomrule
\end{tabular}
\end{table}

\noindent\footnotesize\textit{* The metrics for the marked models are from the VulTriage paper~\cite{tang2026vultriage}.}\normalsize

The upward arrow indicates that a larger value is better, whereas the downward arrow indicates that a smaller value is better. Data preprocessing and the effective test-set size may differ across methods, while ObliVul is evaluated on the 422 test pairs retained after preprocessing using \textbf{Qwen2.5-Coder-7B-Instruct}. The model employs deterministic greedy decoding without stochastic sampling, and no temperature parameter is used. Therefore, the table provides a performance reference rather than a strict same-protocol comparison.

Table~\ref{tab:main-function} shows that traditional pretrained models and some alert-driven methods tend to exhibit two typical problems. One group continues to classify fixed versions as vulnerable and essentially remains at the level of identifying suspicious risk clues. The other group reduces false positives but has weak coverage of actual vulnerabilities. In contrast, ObliVul does not simply increase model sensitivity to suspicious code. It uses LSOG to structurally represent local safety obligations, applies bidirectional counterfactual validation to test whether risk changes are caused by safety obligations, and uses VAFI to aggregate evidence at the function level, thereby achieving a better balance between vulnerability discovery and repair recognition.

On the LinuxVul test set, ObliVul maintains strong performance despite the dataset shift, achieving 89.69\% accuracy, 88.89\% precision, 90.72\% recall, and 89.80\% F1-score. This result indicates good cross-dataset generalization to Linux kernel vulnerability functions.

For the vulnerable/fixed paired data in PrimeVul pair-function, both versions often share substantial context and code structure. Methods relying on surface patterns or alert intensity can easily continue classifying the fixed function as vulnerable. ObliVul uses local semantic differences related to the fix rather than capturing only dangerous operations or code patterns shared by both versions, demonstrating its advantage in repair-semantic modeling.

\begin{table}[!htbp]
\captionsetup{skip=1pt}
\caption{Paired-state diagnosis results.}
\label{tab:main-pair}
\centering
\evaltablesetup
\begin{tabular}{@{}llcccccc@{}}
\toprule
Method & Dataset & Valid pairs & P-C$\uparrow$ & P-V$\downarrow$ & P-B$\downarrow$ & P-R$\downarrow$ & VPS$\uparrow$ \\
\midrule
CodeBERT & PrimeVul & 422 & 0.00\% & 100.00\% & 0.00\% & 0.00\% & 0.00\% \\
CodeT5 & PrimeVul & 422 & 0.00\% & 99.76\% & 0.24\% & 0.00\% & 0.00\% \\
UniXCoder & PrimeVul & 422 & 9.00\% & 42.65\% & 47.16\% & 1.18\% & 7.82\% \\
LineVul & PrimeVul & 422 & 33.89\% & 11.85\% & 51.18\% & 3.08\% & 30.81\% \\
Ding et al. CoT$^{*}$ & PrimeVul & 435 & 9.20\% & 9.89\% & 74.25\% & 6.67\% & 2.53\% \\
GPTLens$^{*}$ & PrimeVul & 435 & 10.11\% & 55.40\% & 28.05\% & 6.44\% & 3.68\% \\
VulTrial$^{*}$ & PrimeVul & 435 & 18.62\% & 41.15\% & 28.74\% & 11.49\% & 7.13\% \\
ConColl$^{*}$ & PrimeVul & 435 & 5.06\% & 77.47\% & 14.94\% & 2.53\% & 2.53\% \\
VulTriage$^{*}$ & PrimeVul & 435 & 33.79\% & 47.59\% & 11.95\% & 6.67\% & 27.13\% \\
CLEAR~\cite{yun2026clear} & PrimeVul & Not reported & 19.17\% & 4.62\% & 67.67\% & 8.55\% & 10.62\% \\
\textbf{ObliVul} & \textbf{PrimeVul} & \textbf{422} & \textbf{59.72\%} & \textbf{27.73\%} & \textbf{0.00\%} & \textbf{12.56\%} & \textbf{47.16\%} \\
\textbf{ObliVul} & \textbf{LinuxVul} & \textbf{97} & \textbf{79.38\%} & \textbf{11.34\%} & \textbf{9.28\%} & \textbf{0.00\%} & \textbf{79.38\%} \\
\bottomrule
\end{tabular}
\end{table}

\noindent\footnotesize\textit{* The metrics for the marked models are from the VulTriage paper~\cite{tang2026vultriage}.}\normalsize

The number of valid pairs differs across methods. Each percentage uses the number of pairs actually evaluated by the corresponding method as its denominator, and CLEAR does not report the number of valid pairs. The results describe prediction behavior within each method's own test scope and do not constitute a strict same-protocol comparison.

Table~\ref{tab:main-pair} further reveals the essential differences among methods from the paired-state perspective. Many methods tend to classify both vulnerable and fixed versions as vulnerable, indicating that they primarily learn risk clues such as dangerous operations, sensitive APIs, or code patterns but do not further determine whether the conditions required for a vulnerability remain satisfied.

ObliVul's advantage comes from structured modeling of vulnerability-repair semantics. LSOG explicitly represents safety obligations and their scopes, bidirectional counterfactual validation examines whether changes in safety obligations cause risk responses, and VAFI aggregates verified alert evidence into a function-level judgment. Together, the three components enable the model to perform both vulnerability recognition and repair recognition, avoiding persistent alerts on code that has been fixed or is constrained by effective safety obligations.

\subsection{Ablation Experiments}
This section conducts ablation experiments on the three core modules of ObliVul using the PrimeVul dataset described in Section~4.1.1 to analyze the contribution of each component. It addresses the following research questions:

\textbf{RQ1:} Can LSOG provide effective vulnerability-semantic information?

\textbf{RQ2:} Is bidirectional counterfactual validation better than no validation or one-directional validation?

\textbf{RQ3:} Can VAFI aggregation and consistency learning improve function-level vulnerability judgment?

All ablation experiments use the same fixed data split and test protocol. Model selection, hyperparameter selection, and threshold selection are performed using the validation set only, and the test set is used only for final evaluation. For all three ablation tables, P-C, P-V, P-B, P-R, and VPS are calculated over the 422 valid sample pairs and reported as percentages.

\subsubsection{RQ1: Effect of LSOG-Based Structured Semantics}
This experiment compares a base variant using CPG, LEP, and alert features; an LSOG variant containing core security entities; a variant using the complete LSOG and validation-aware features; and a variant without structured semantic evidence. Table~\ref{tab:lsog-ablation} presents the results.

\begin{table}[!htbp]
\captionsetup{skip=1pt}
\caption{LSOG structured-semantics ablation.}
\label{tab:lsog-ablation}
\centering
\evaltablesetup
\setlength{\tabcolsep}{0.75pt}
\begin{tabular}{@{}p{0.34\linewidth}*{9}{c}@{}}
\toprule
Variant & Acc$\uparrow$ & Pre$\uparrow$ & Recall$\uparrow$ & F1$\uparrow$ & P-C$\uparrow$ & P-V$\downarrow$ & P-B$\downarrow$ & P-R$\downarrow$ & VPS$\uparrow$ \\
\midrule
A0: w/o Structured Semantics & 0.5201 & 0.5107 & 0.9645 & 0.6678 & 7.58\% & 88.86\% & 0.00\% & 3.55\% & 4.03\% \\
A1: CPG + LEP + Alert & 0.6315 & 0.5920 & 0.8460 & 0.6966 & 41.71\% & 42.89\% & 0.00\% & 15.40\% & 26.30\% \\
A2: Core LSOG & 0.6374 & 0.5973 & 0.8436 & 0.6994 & 43.13\% & 41.23\% & 0.00\% & 15.64\% & 27.49\% \\
A3: Full LSOG + Validation-Aware & 0.7358 & 0.6846 & 0.8744 & 0.7680 & 59.72\% & 27.73\% & 0.00\% & 12.56\% & 47.16\% \\
\bottomrule
\end{tabular}
\end{table}

Table~\ref{tab:lsog-ablation} shows that the value of LSOG does not lie in simply adding several security-entity labels. Instead, it organizes scattered local program facts into a structured semantic representation consisting of risk entities, protection conditions, scopes, and dependencies. Adding only core entities such as source, sink, guard, and sanitizer is insufficient to explain whether a vulnerability holds. The complete LSOG further represents whether protection conditions are effective, whether state preconditions are satisfied, whether critical dependencies cover the dangerous path, and the scope of each safety obligation. It therefore enables the model to progress from detecting a dangerous operation in code to determining whether that operation constitutes a real vulnerability in the current context.

\subsubsection{RQ2: Effect of Bidirectional Counterfactual Validation}
This experiment compares five settings: no counterfactual validation, forward-only validation, direct bidirectional validation using both forward and reverse risk responses, bidirectional validation without minimal supporting evidence, and complete bidirectional validation with minimal supporting evidence and validation-quality fusion. Table~\ref{tab:cf-ablation} presents the results.

\begin{table}[!htbp]
\captionsetup{skip=1pt}
\caption{Bidirectional-validation ablation.}
\label{tab:cf-ablation}
\centering
\evaltablesetup
\setlength{\tabcolsep}{0.75pt}
\begin{tabular}{@{}p{0.34\linewidth}*{9}{c}@{}}
\toprule
Variant & Acc$\uparrow$ & Pre$\uparrow$ & Recall$\uparrow$ & F1$\uparrow$ & P-C$\uparrow$ & P-V$\downarrow$ & P-B$\downarrow$ & P-R$\downarrow$ & VPS$\uparrow$ \\
\midrule
B0: w/o Counterfactual Validation & 0.5569 & 0.5311 & 0.9716 & 0.6868 & 14.22\% & 82.94\% & 0.00\% & 2.84\% & 11.37\% \\
B1: Forward-Only Validation & 0.5379 & 0.5199 & 0.9905 & 0.6819 & 8.53\% & 90.52\% & 0.00\% & 0.95\% & 7.58\% \\
B2: Direct Bidirectional Validation & 0.5415 & 0.5218 & 0.9929 & 0.6841 & 9.00\% & 90.28\% & 0.00\% & 0.71\% & 8.29\% \\
B3: Bidirectional w/o Minimal Evidence & 0.6422 & 0.6003 & 0.8507 & 0.7039 & 43.36\% & 41.71\% & 0.00\% & 14.93\% & 28.44\% \\
B4: Full Bidirectional Validation & 0.7287 & 0.6764 & 0.8768 & 0.7637 & 58.06\% & 29.62\% & 0.00\% & 12.32\% & 45.73\% \\
\bottomrule
\end{tabular}
\end{table}

Table~\ref{tab:cf-ablation} shows that merely introducing counterfactual risk responses does not automatically improve vulnerability discrimination. B1 uses only the forward intervention, whereas B2 directly combines the forward and reverse responses while masking witness/O*, risk-flip, pair-prior, source-sink, path/slice, and patch-quality evidence. Consequently, both variants retain high recall but continue to classify many fixed versions as vulnerable.

B3 reconstructs the bidirectional validation-aware representation but removes minimal supporting evidence. It substantially improves precision, accuracy, and paired-state discrimination. B4 further incorporates witness/O*, contradiction information, risk-flip evidence, and validation-quality fusion. This produces the strongest controlled RQ2 result, demonstrating that the benefit comes not only from observing risk changes in two directions, but also from verifying whether those changes are supported by consistent local program evidence.

The comparison result in Tables~\ref{tab:main-function} and~\ref{tab:main-pair} corresponds to the validation-selected deployment configuration. Table~\ref{tab:cf-ablation} reports the controlled RQ2 branches and separately evaluates the complete minimal-evidence-enhanced counterfactual mechanism.

\subsubsection{RQ3: Effect of VAFI Aggregation and Consistency Learning}
This experiment compares Mean Pooling, Attention Pooling, base VAFI, variants that separately add ranking, path-alignment, and hierarchical-consistency constraints, and complete consistency learning. Table~\ref{tab:vafi-ablation} presents the results.

\begin{table}[!htbp]
\captionsetup{skip=1pt}
\caption{VAFI aggregation and consistency ablation.}
\label{tab:vafi-ablation}
\centering
\evaltablesetup
\setlength{\tabcolsep}{0.75pt}
\begin{tabular}{@{}p{0.34\linewidth}*{9}{c}@{}}
\toprule
Variant & Acc$\uparrow$ & Pre$\uparrow$ & Recall$\uparrow$ & F1$\uparrow$ & P-C$\uparrow$ & P-V$\downarrow$ & P-B$\downarrow$ & P-R$\downarrow$ & VPS$\uparrow$ \\
\midrule
C0: Mean Pooling & 0.6611 & 0.6223 & 0.8199 & 0.7076 & 45.50\% & 36.49\% & 4.74\% & 13.27\% & 32.23\% \\
C1: Attention Pooling & 0.6801 & 0.6352 & 0.8460 & 0.7256 & 49.05\% & 35.55\% & 2.37\% & 13.03\% & 36.02\% \\
C2: Base VAFI & 0.6872 & 0.6367 & 0.8720 & 0.7360 & 47.39\% & 39.81\% & 2.84\% & 9.95\% & 37.44\% \\
C3: C2 + $\mathcal{L}_{rank}$ & 0.6777 & 0.6298 & 0.8626 & 0.7280 & 47.16\% & 39.10\% & 2.13\% & 11.61\% & 35.55\% \\
C4: C2 + $\mathcal{L}_{align}$ & 0.6848 & 0.6383 & 0.8531 & 0.7302 & 49.05\% & 36.26\% & 2.61\% & 12.09\% & 36.97\% \\
C5: C2 + $\mathcal{L}_{cons}$ & 0.6860 & 0.6365 & 0.8673 & 0.7342 & 44.55\% & 42.18\% & 5.92\% & 7.35\% & 37.20\% \\
C6: Complete VAFI & 0.7358 & 0.6846 & 0.8744 & 0.7680 & 59.72\% & 27.73\% & 0.00\% & 12.56\% & 47.16\% \\
\bottomrule
\end{tabular}
\end{table}

Table~\ref{tab:vafi-ablation} shows that function-level vulnerability judgment cannot be reduced to simple averaging or weighted aggregation of alert representations or alert scores. Simple aggregation does not explicitly distinguish complementarity, redundancy, or validation reliability among alerts. It can repeatedly amplify local risk from the same dangerous path and has difficulty reflecting the combined evidential value of alerts from different paths.

\subsubsection{Summary of the Ablation Experiments}
In summary, LSOG organizes local program facts as structured safety obligations, bidirectional counterfactual validation examines the response relationship between safety obligations and alert risk, and VAFI aggregates verified alert evidence into a function-level vulnerability conclusion.

\renewcommand{\thetable}{\arabic{table}}
\section{Behavioral and Trustworthiness Analysis}
\label{sec:behavior}

\subsection{Reliability Analysis of Risk Outputs}

\subsubsection{Probability Calibration Analysis}

This subsection analyzes the reliability of risk outputs using reliability diagrams, Expected Calibration Error (ECE), and the Brier Score~\cite{guo2017calibration,brier1950}. ECE measures the weighted average deviation across the intervals, whereas the Brier Score measures the overall probabilistic error between the predicted scores and the ground-truth labels. To calculate ECE, the interval $[0,1]$ is equally divided into ten bins, and only non-empty bins are weighted according to their proportions of samples. To compare the states before and after calibration, a temperature-scaling parameter is fitted on the validation set and then fixed and applied to the test set. Test-set labels are not used to fit the calibrator.

This section separately analyzes the raw model outputs and the pair-adjusted outputs. The former are the risk scores directly produced by the random forest, whereas the latter combine the respective outputs of vulnerable and fixed versions to perform paired risk adjustment and are used to examine score reliability under paired analysis.

\begin{figure}[!t]
\centering
\includegraphics[width=0.855\linewidth]{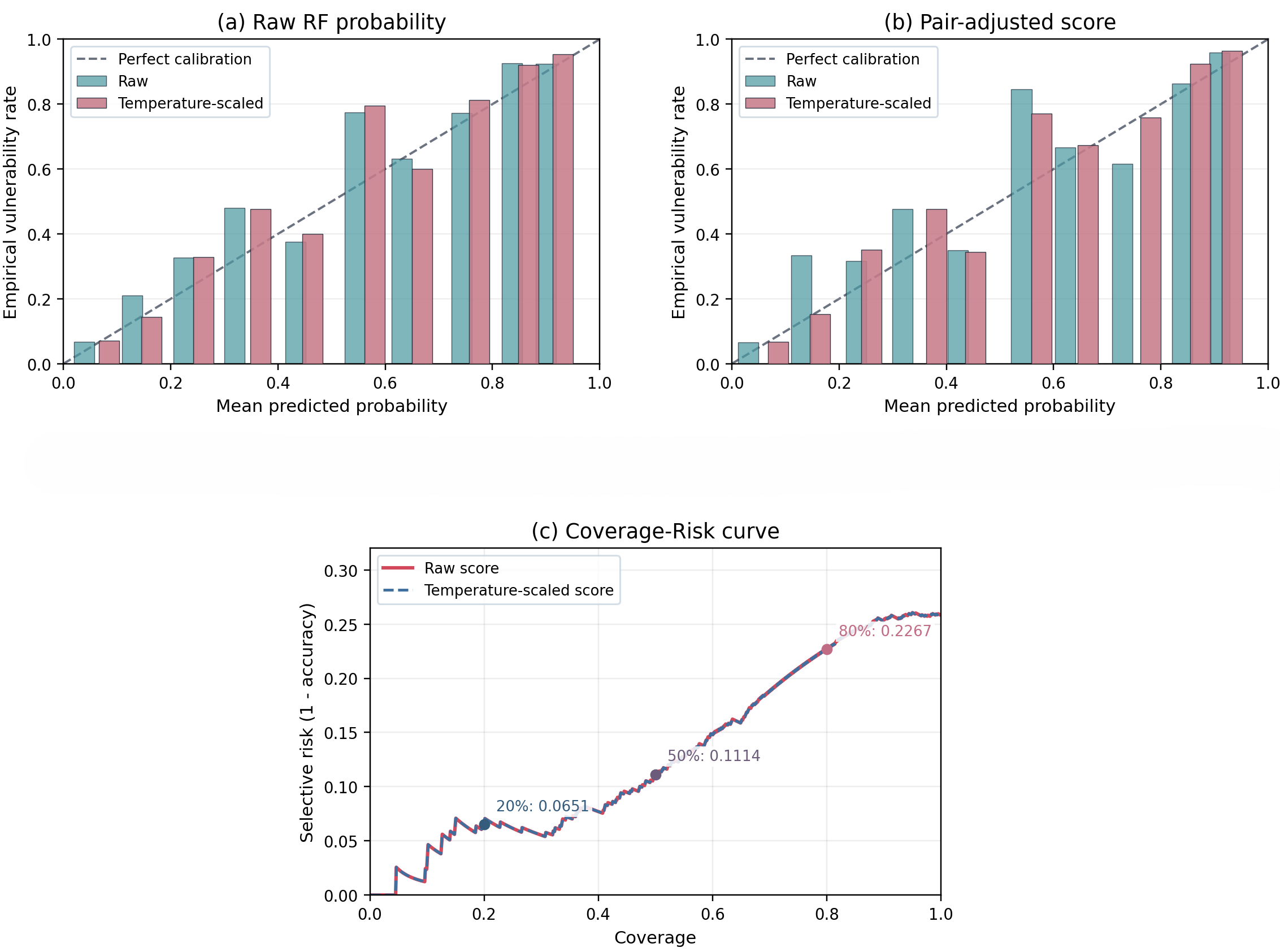}
\caption{Calibration and selective-prediction analysis of function-level risk outputs: (a) reliability of raw model scores before and after calibration; (b) reliability of pair-adjusted scores before and after calibration; and (c) Coverage--Risk curves of pair-adjusted scores before and after calibration.}
\Description{Three panels show the reliability of raw and pair-adjusted risk scores before and after calibration and the Coverage--Risk curves of pair-adjusted scores.}
\label{fig:calibration-selective}
\end{figure}

Figure~\ref{fig:calibration-selective}(a) shows that the raw scores generally correspond to the actual vulnerability proportions and that temperature scaling mainly makes minor corrections to local deviations. Figure~\ref{fig:calibration-selective}(b) shows that paired risk adjustment does not automatically improve probability calibration, but overall reliability does not decrease substantially, and the probabilistic deviations can be further corrected through post-processing fitted on the validation set.

\subsubsection{Selective Prediction Analysis}

The previous subsection discusses the correspondence between risk scores and actual vulnerability proportions. In practical vulnerability review, however, it is also necessary to determine which predictions should be prioritized. We therefore use Coverage--Risk curves to analyze the selective-prediction capability of the model~\cite{geifman2019selectivenet}.

In this experiment, the pair-adjusted function-level risk scores are used to rank the test functions in descending order of prediction confidence, $\max(p_F,1-p_F)$. For a given Coverage, the corresponding proportion of top-ranked functions is retained, and their prediction error rate under the classification threshold of 0.5 is calculated as Risk. As Coverage increases, the functions included in the calculation gradually expand from high-confidence predictions to low-confidence predictions. The resulting Coverage--Risk curve is used to examine whether prediction confidence can distinguish function sets with different error rates.

Figure~\ref{fig:calibration-selective}(c) shows the Coverage--Risk curves of the pair-adjusted scores before and after calibration. As the coverage range expands, the prediction error rate generally increases, indicating that the model can use prediction confidence to concentrate relatively reliable judgments at the top of the ranking. The function-level output of ObliVul not only provides vulnerability classes but also supplies a continuous signal that distinguishes different levels of prediction reliability. Security analysts can therefore rapidly inspect high-confidence predictions, concentrate more detailed manual review on low-confidence predictions, and combine the alert-level evidence provided by VAFI to further locate the local paths requiring examination.

\subsection{Paired Counterfactual Risk-Response Analysis}

Classification metrics show whether functions are correctly identified but not whether models capture security-state changes caused by fixes. In PrimeVul pair-function, each vulnerable/fixed pair comes from the same fix and shares code structure and functional semantics; differences concern local safety obligations such as boundary checks, state constraints, and data sanitization. We compare pre- and post-fix risks and define the risk difference of pair $i$ as follows:

\begin{equation}
\Delta r_i=R(x_i^{vul})-R(x_i^{fix}).
\label{eq:paired-risk-difference}
\end{equation}

Here, $R(x_i^{vul})$ and $R(x_i^{fix})$ denote the risk scores of the vulnerable and fixed versions, respectively. When $\Delta r_i>0$, the direction of risk change is consistent with the actual repair relationship. When $\Delta r_i<0$, the risk ordering of the two versions is reversed. When $\Delta r_i=0$, the model fails to identify the risk difference before and after repair. This metric reflects whether the model can adjust its risk judgment according to the addition, removal, or state changes of safety obligations. Table~\ref{tab:risk-response} presents the experimental results.

\begin{table}[!t]
\captionsetup{skip=1pt}
\caption{Counterfactual risk-response results of the main model on paired samples.}
\label{tab:risk-response}
\centering
\fontsize{8.2pt}{9.4pt}\selectfont
\setlength{\tabcolsep}{3pt}
\renewcommand{\arraystretch}{1.00}
\begin{tabularx}{\linewidth}{@{}>{\raggedright\arraybackslash}p{0.31\linewidth}>{\centering\arraybackslash}p{0.19\linewidth}X@{}}
\toprule
Evaluation metric & Result & Interpretation \\
\midrule
Mean vulnerable-version risk $R_{vul}$ & 0.6120 & Average risk level before repair \\
Mean fixed-version risk $R_{fix}$ & 0.2511 & Average risk level after repair \\
Mean risk difference $\Delta r$ & +0.3609 & Average risk reduction after repair \\
Median risk difference $\Delta r$ & +0.4545 & Risk reduction of a typical sample \\
Positive response ($\Delta r>0$) & 274/422 (64.9\%) & Correctly identifies the risk reduction caused by repair \\
Reverse response ($\Delta r<0$) & 71/422 (16.8\%) & Risk increases after repair \\
No response ($\Delta r=0$) & 77/422 (18.2\%) & Fails to identify the risk change before and after repair \\
Pair-Correct & 252/422 (59.7\%) & Both versions are correctly classified \\
Risk Flip & 253/422 (60.0\%) & Risk scores cross the 0.5 risk boundary \\
Wilcoxon signed-rank test~\cite{wilcoxon1945} & $W=52132$, $p<0.001$ & Overall risk differences are significantly greater than zero \\
Rank-biserial correlation $r_{rb}$ & 0.747 & Strong risk separation before and after repair \\
\bottomrule
\end{tabularx}
\end{table}

\noindent\textit{Note:} Pair-Correct is calculated using the classification threshold of 0.32 determined on the validation set. Risk Flip uses 0.5 as the midpoint of the risk score and requires the vulnerable-version risk to be no lower than 0.5 and the fixed-version risk to be lower than 0.5.

Table~\ref{tab:risk-response} shows that repair operations cause directionally correct and practically meaningful risk changes in most samples. Paired comparison controls differences in function structure, functional semantics, and complexity, indicating that ObliVul can identify local repair information and transform changes in safety obligations, including boundary checks, state constraints, and data sanitization, into stable risk adjustments.

\begin{figure}[!t]
\centering
\includegraphics[width=\linewidth]{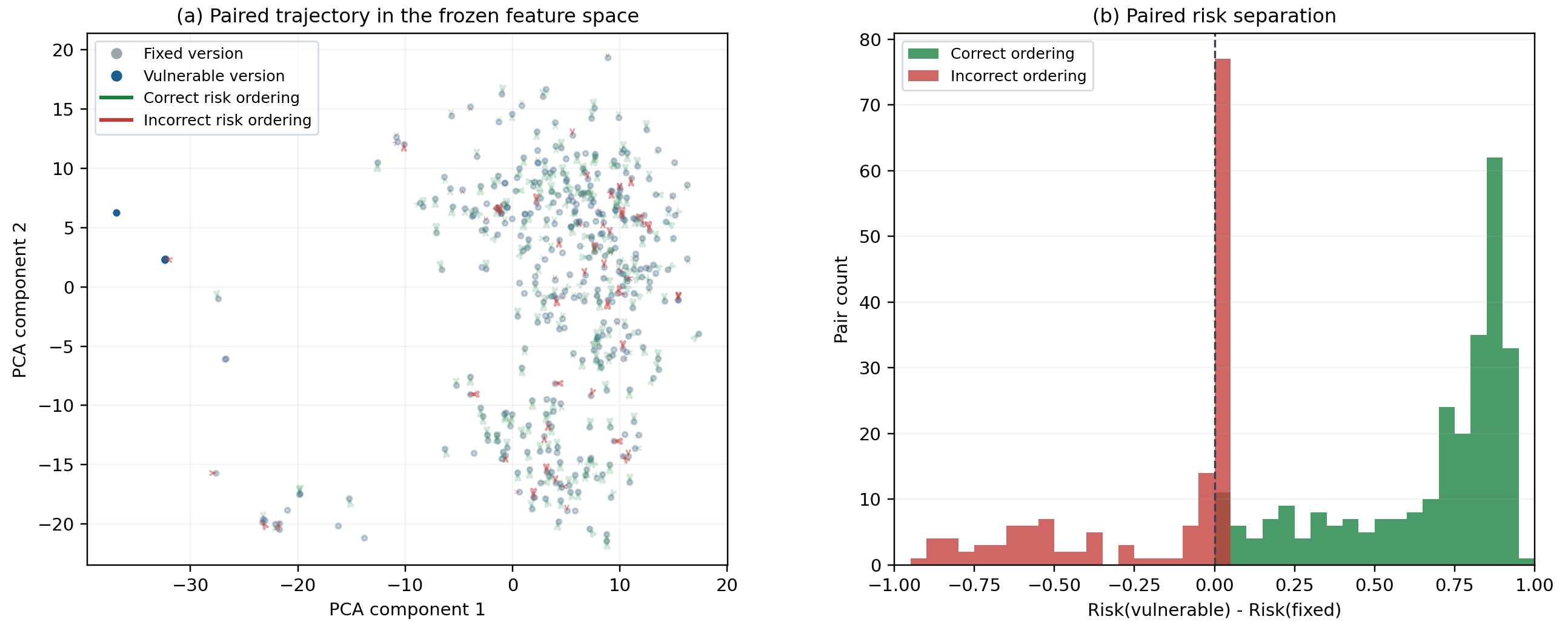}
\caption{Paired risk response of ObliVul: (a) feature trajectories directed from fixed versions to vulnerable versions; and (b) distribution of paired risk differences.}
\Description{The left panel shows feature-space trajectories from fixed versions to vulnerable versions, and the right panel shows the distribution of paired risk differences.}
\label{fig:paired-risk}
\end{figure}

Figure~\ref{fig:paired-risk}(a) projects the frozen fused features into a two-dimensional space. Each arrow points from a fixed version to its corresponding vulnerable version. Green indicates $R(x_{vul})>R(x_{fix})$, whereas red indicates the opposite risk ordering. Although the two versions do not form completely separated clusters, most trajectories move in the direction of increasing risk. Figure~\ref{fig:paired-risk}(b) further shows that the risk differences are mainly distributed to the right of zero. Together, the two panels demonstrate that ObliVul can encode risk differences caused by changes in local safety obligations while preserving the main structures and functional semantics shared before and after repair.

In summary, ObliVul can not only identify static dangerous patterns but also capture repair-related security semantics when code structures and functional semantics are highly similar, thereby enabling fine-grained judgments of changes in safety-obligation states. Compared with methods that perform classification based only on surface-level code features, ObliVul exhibits stronger sensitivity to repair semantics and better risk-discrimination capability.

\vspace{-4pt}
\subsection{Preliminary Validation on a Real Linux Kernel Project}

To assess ObliVul beyond public datasets, we apply the trained model to Linux kernel vulnerability screening: alerts are generated within a specified scope, high-review-value locations are selected by top-$k$ filtering, and candidate risks are checked using manual review and upstream repair information. Three identified real defects have been fixed, with their patches reviewed and merged into official Linux commits; one other concurrently submitted candidate and its corresponding patch remain under validation because they require time-consuming physical-machine PCI tests. (Links to the corresponding Linux commits will be made publicly available upon publication.) These results preliminarily show that, in real projects, ObliVul extracts alert-centered local evidence, relates data propagation, boundary constraints, control conditions, and protection mechanisms, and combines alerts, local evidence, and pre-/post-fix differences to narrow manual audits, locate critical program positions, and provide structured evidence for understanding risk formation.

\vspace{-4pt}
\section{Conclusion and Future Work}
\label{sec:conclusion}

To address incomplete alert contexts, high false-positive rates, and the difficulty of explaining why vulnerabilities hold, this paper proposes ObliVul, an alert-driven bidirectional counterfactual graph-validation framework. ObliVul extracts local program evidence around an individual alert, uses a large language model to recover safety obligations and construct a Local Safety Obligation Graph, validates risk responses by adding missing protection and removing existing protection, and finally uses VAFI to aggregate alert evidence and produce a function-level vulnerability judgment. Experiments on PrimeVul pair-function show that ObliVul can reduce persistent false positives on fixed versions and more stably distinguish vulnerable versions from fixed versions. Probability calibration, selective prediction, and validation on a real-world Linux kernel project further show that ObliVul can provide risk scores, audit priorities, and local evidence for review, thereby improving the interpretability and practical engineering value of the detection results.

ObliVul continues to depend on the semantic-recovery capability of large language models and the quality of local CPGs. Future work will combine symbolic execution, dynamic testing, and runtime information to improve complex-path and environmental analysis.

\paragraph{Generative AI Disclosure.}
Generative AI tools were used only to assist the visual presentation of Figures~1 and~2; their technical content, labels, and interpretations were designed and verified by the authors.

\end{document}